\documentclass[sigconf]{acmart}

\usepackage{booktabs} 
\usepackage{siunitx}  

\AtBeginDocument{%
	\providecommand\BibTeX{{%
		\normalfont B\kern-0.5em{\scshape i\kern-0.25em b}\kern-0.8em\TeX}}}

\copyrightyear{2024}
\acmYear{2024}
\setcopyright{acmlicensed}
\acmConference[SIGIR '24]{Proceedings of the 47th International ACM SIGIR Conference on Research and Development in Information Retrieval}{July 14--18, 2024}{Washington, DC, USA}
\acmBooktitle{Proceedings of the 47th International ACM SIGIR Conference on Research and Development in Information Retrieval (SIGIR '24), July 14--18, 2024, Washington, DC, USA}
\acmDOI{10.1145/3626772.3661370}
\acmISBN{979-8-4007-0431-4/24/07}
\begin{document}


 	\title{Retrieval-Augmented Generation with Knowledge Graphs for Customer Service Question Answering}

	\author{Zhentao Xu}
	\orcid{0009-0000-5827-7879}
	\email{zhexu@linkedin.com}
	\affiliation{%
		\institution{LinkedIn Corporation}
		\streetaddress{1000 W Maude Ave}
		\city{Sunnyvale}
		\state{CA}
		\country{USA}
		\postcode{94089}
	}

	\author{Mark Jerome Cruz}
        \orcid{0009-0009-4121-6880}
	\email{marcruz@linkedin.com}
	\affiliation{%
		\institution{LinkedIn Corporation}
		\streetaddress{1000 W Maude Ave}
		\city{Sunnyvale}
		\state{CA}
		\country{USA}
		\postcode{94089}
	}

	\author{Matthew Guevara}
        \orcid{0009-0009-1570-3730}
	\email{mguevara@linkedin.com}
	\affiliation{%
		\institution{LinkedIn Corporation}
		\streetaddress{1000 W Maude Ave}
		\city{Sunnyvale}
		\state{CA}
		\country{USA}
		\postcode{94089}
	}

	\author{Tie Wang}
        \orcid{0009-0009-5345-1773}
	\email{tiewang@linkedin.com}
	\affiliation{%
		\institution{LinkedIn Corporation}
		\streetaddress{1000 W Maude Ave}
		\city{Sunnyvale}
		\state{CA}
		\country{USA}
		\postcode{94089}
	}

	\author{Manasi Deshpande}
        \orcid{0009-0007-5670-2748}
	\email{madeshpande@linkedin.com}
	\affiliation{%
		\institution{LinkedIn Corporation}
		\streetaddress{1000 W Maude Ave}
		\city{Sunnyvale}
		\state{CA}
		\country{USA}
		\postcode{94089}
	}

	\author{Xiaofeng Wang}
        \orcid{0009-0009-5648-4183}
	\email{xiaofwang@linkedin.com}
	\affiliation{%
		\institution{LinkedIn Corporation}
		\streetaddress{1000 W Maude Ave}
		\city{Sunnyvale}
		\state{CA}
		\country{USA}
		\postcode{94089}
	}

	\author{Zheng Li}
        \orcid{0009-0006-6892-3073}
	\email{zeli@linkedin.com}
	\affiliation{%
		\institution{LinkedIn Corporation}
		\streetaddress{1000 W Maude Ave}
		\city{Sunnyvale}
		\state{CA}
		\country{USA}
		\postcode{94089}
	}

\renewcommand{\shortauthors}{Zhentao Xu, et al.}

\begin{abstract}
In customer service technical support, swiftly and accurately retrieving relevant past issues is critical for efficiently resolving customer inquiries. The conventional retrieval methods in retrieval-augmented generation (RAG) for large language models (LLMs) treat a large corpus of past issue tracking tickets as plain text, ignoring the crucial intra-issue structure and inter-issue relations, which limits performance. We introduce a novel customer service question-answering method that amalgamates RAG with a knowledge graph (KG). Our method constructs a KG from historical issues for use in retrieval, retaining the intra-issue structure and inter-issue relations. During the question-answering phase, our method parses consumer queries and retrieves related sub-graphs from the KG to generate answers. This integration of a KG not only improves retrieval accuracy by preserving customer service structure information but also enhances answering quality by mitigating the effects of text segmentation. Empirical assessments on our benchmark datasets, utilizing key retrieval (MRR, Recall@K, NDCG@K) and text generation (BLEU, ROUGE, METEOR) metrics, reveal that our method outperforms the baseline by 77.6\% in MRR and by 0.32 in BLEU. Our method has been deployed within LinkedIn’s customer service team for approximately six months and has reduced the median per-issue resolution time by 28.6\%.
\end{abstract}



	\maketitle

	\section{Introduction}

Effective technical support in customer service underpins product success, directly influencing customer satisfaction and loyalty. Given the frequent similarity of customer inquiries to previously resolved issues, the rapid and accurate retrieval of relevant past instances is crucial for the efficient resolution of such inquiries. Recent advancements in embedding-based retrieval (EBR), large language models (LLMs), and retrieval-augmented generation (RAG) \cite{lewis2020retrieval} have significantly enhanced retrieval performance and question-answering capabilities for the technical support of customer service. This process typically unfolds in two stages: first, historical issue tickets are treated as plain text, segmented into smaller chunks to accommodate the context length constraints of embedding models; each chunk is then converted into an embedding vector for retrieval. Second, during the question-answering phase, the system retrieves the most relevant chunks and feeds them as contexts for LLMs to generate answers in response to queries. Despite its straightforward approach, this method encounters several limitations:

\begin{itemize}
    \item \textbf{Limitation 1 - Compromised Retrieval Accuracy from Ignoring Structures:} Issue tracking documents such as Jira \cite{AtlassianJira} possess inherent structure and are interconnected, with references such as "issue A is related to/copied from/caused by issue B." The conventional approach of compressing documents into text chunks leads to the loss of vital information. Our approach parses issue tickets into trees and further connects individual issue tickets to form an interconnected graph, which maintains this intrinsic relationship among entities, achieving high retrieval performance.
    \item \textbf{Limitation 2 - Reduced Answer Quality from Segmentation:} Segmenting extensive issue tickets into fixed-length segments to accommodate the context length constraints of embedding models can result in the disconnection of related content, leading to incomplete answers. For example, an issue ticket describing an issue at its beginning and its solution at the end may be split during the text segmentation process, resulting in the omission of critical parts of the solution. Our graph-based parsing method overcomes this by preserving the logical coherence of ticket sections, ensuring the delivery of complete and high-quality responses.
\end{itemize}

\section{Related Work}

Question answering (QA) with knowledge graphs (KGs) can be broadly classified into retrieval-based, template-based, and semantic parsing-based methods. Retrieval-based approaches utilize relation extraction \cite{xu2016hybrid} or distributed representations \cite{bordes2014open} to derive answers from KGs, but they face difficulties with questions involving multiple entities. Template-based strategies depend on manually-created templates for encoding complex queries, yet are limited by the scope of available templates \cite{unger2012template}. Semantic parsing methods map text to logical forms containing predicates from KGs \cite{bhutani2020answering}  \cite{sun-etal-2018-open} \cite{yih2015semantic}.

Recent advancements in large language models (LLMs) integration with Knowledge Graphs (KGs) have demonstrated notable progress. Jin et al. \cite{jin2023large} provide a comprehensive review of this integration, categorizing the roles of LLMs as Predictors, Encoders, and Aligners. For graph-based reasoning, Think-on-Graph \cite{sun2023think} and Reasoning-on-Graph \cite{luo2023reasoning} enhance LLMs' reasoning abilities by integrating KGs. Yang et al. \cite{yang2023chatgpt} propose augmenting LLMs' factual reasoning across various training phases using KGs. For LLM-based question answering, Wen et al.'s Mindmap \cite{wen2023mindmap} and Qi et al. \cite{qi2023foodgpt} employ KGs to boost LLM inference capabilities in specialized domains such as medicine and food. These contributions underscore the increasing efficacy of LLM and KG combinations in enhancing information retrieval and reasoning tasks.

\section{Methods}

\begin{figure*}[t] 
        \centering
        \includegraphics[width=\textwidth]{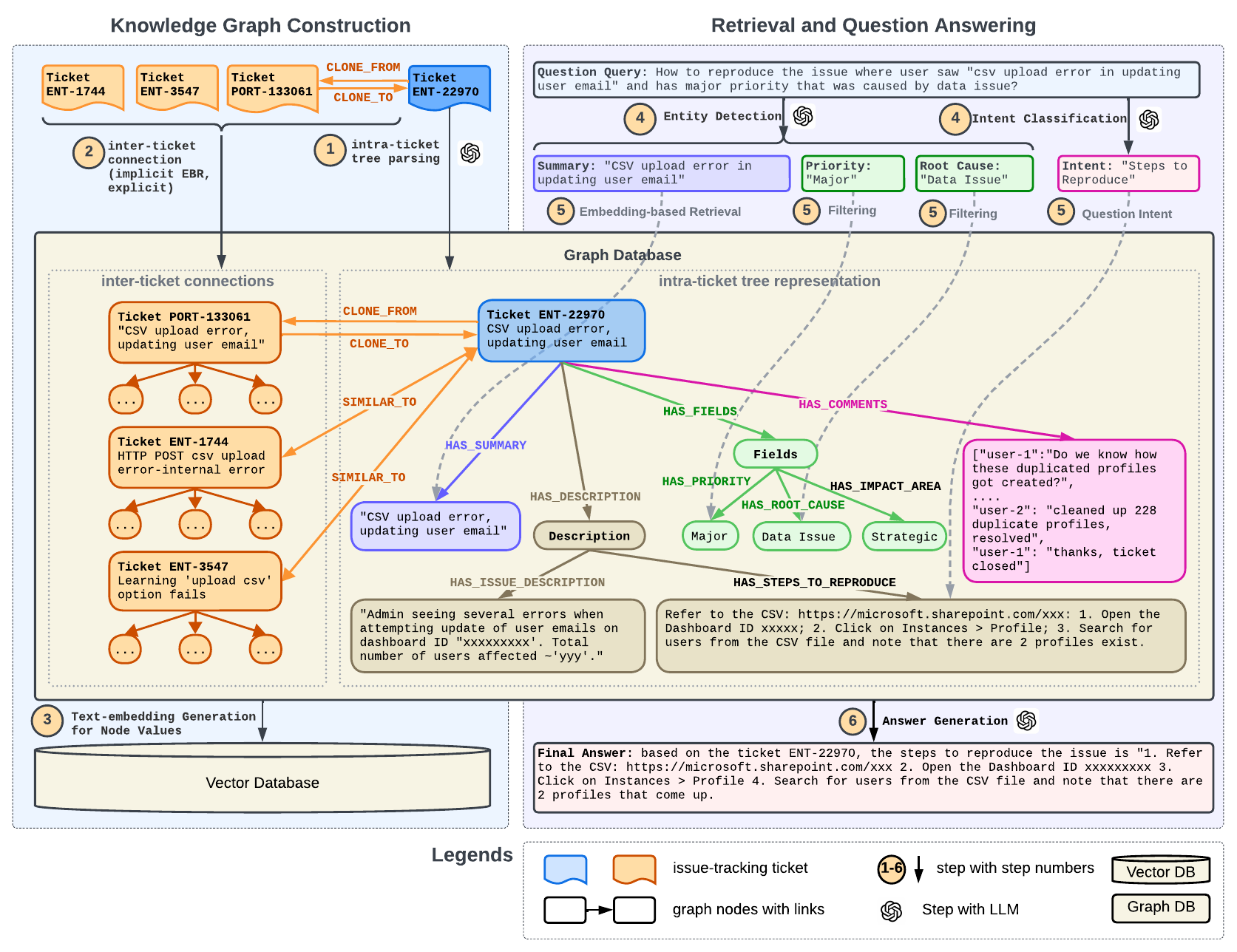}
        \caption{An overview of our proposed retrieval-augmented generation with knowledge graph framework. The left side of this diagram illustrates the knowledge graph construction; the right side shows the retrieval and question answering process.}
        \label{fig:geebr_system_diagram}
    \end{figure*}

We introduce an LLM-based customer service question answering system that seamlessly integrates retrieval-augmented generation (RAG) with a knowledge graph (KG). Our system (Figure \ref{fig:geebr_system_diagram}) comprises two phases: First, during the KG construction phase, our system constructs a comprehensive knowledge graph from historical customer service issue tickets. It integrates a tree-structured representation of each issue and interlinks them based on relational context. It also generates embedding for each node to facilitate later semantic searching. Second, during the question-answering phase, our method parses consumer queries to identify named entities and intents. It then navigates within the KG to identify related sub-graphs for generating answers.

\subsection{Knowledge Graph Construction}

\subsubsection{Graph Structure Definition}
In defining the knowledge graph structure for historical issue representation, we employ a dual-level architecture that segregates intra-issue and inter-issue relations, as illustrated in Figure \ref{fig:geebr_system_diagram}. The \textbf{Intra-issue Tree} \(\mathcal{T}_i(\mathcal{N}, \mathcal{E}, \mathcal{R})\) models each ticket \(t_i\) as a tree, where each node \(n \in \mathcal{N}\), identified by a unique combination \((i, s)\), corresponds to a distinct section \(s\) of ticket \(t_i\), and each edge \(e \in \mathcal{E}\) and \(r \in \mathcal{R}\) signifies the hierarchical connection and type of relations between these sections. The \textbf{Inter-issue Graph} \(\mathcal{G}(\mathcal{T}, \mathcal{E}, \mathcal{R})\) represents the network of connections across different tickets, incorporating both explicit links \(\mathcal{E}_{\text{exp}}\), defined in issue tracking tickets, and implicit connections \(\mathcal{E}_{\text{imp}}\), derived from semantic similarity between tickets. For implicit connections, we leverage cosine similarity between the embedding vectors of ticket titles, a method adaptable to specific use cases.

For instance, Figure \ref{fig:geebr_system_diagram} portrays ticket \texttt{ENT-22970} as a tree structure with nodes representing sections such as \texttt{Summary}, \texttt{Description}, and \texttt{Priority}. It exhibits a direct \texttt{clone} linkage to \texttt{PORT-133061}, indicating an explicit \texttt{clone} relationship. Additionally, it’s implicitly connected with \texttt{ENT-1744} and \texttt{ENT-3547} due to the semantic similarities.

\subsubsection{Knowledge Graph Construction}
\label{subsubsection: offline_graph_construction}

Graph construction is delineated into two phases: intra-ticket parsing and inter-ticket connection. \textbf{1) Intra-Ticket Parsing Phase:} This phase transforms each text-based ticket \(t_i\) into a tree representation \( \mathcal{T}_i \). We employ a hybrid methodology, initially utilizing rule-based extraction for predefined fields, such as code sections identified via keywords. Subsequently, for text not amenable to rule-based parsing, we engage an LLM for parsing. The LLM is directed by a YAML template \( \mathcal{T}_{\text{template}} \), representing in graph the ticket sections routinely utilized by customer support. \textbf{2) Inter-Ticket Connection Phase:} Here, individual trees \( \mathcal{T}_i \) are amalgamated into a comprehensive graph \( \mathcal{G} \). Explicit connections \( \mathcal{E}_{\text{exp}} \) are delineated as specified within tickets, exemplified by designated fields in Jira \cite{AtlassianJira}. Implicit connections \( \mathcal{E}_{\text{imp}} \) are inferred from textual-semantic similarities across ticket titles, employing embedding techniques and a threshold mechanism to discern the most relevant tickets for each issue ticket.
$$ t_i = t_{i, \text{rule}} \cup t_{i, \text{llm}} $$
$$ \mathcal{T}_i = \text{RuleParse}(t_{i, \text{rule}}) + \text{LLMParse}(t_{i, \text{llm}}, \mathcal{T}_{\text{template}}, \text{prompt}) $$
$$ \mathcal{E}_{\text{exp}} = \{(\mathcal{T}_i, \mathcal{T}_j) \mid \mathcal{T}_i \text{ explicitly connected to } \mathcal{T}_j\} $$
$$ \mathcal{E}_{\text{imp}} = \{(\mathcal{T}_i, \mathcal{T}_j) \mid \text{cos}(\text{embed}(\mathcal{T}_i), \text{embed}(\mathcal{T}_j)) \geq \theta \}$$

\subsubsection{Embedding Generation}
\label{subsubsection: embedding_generation}
To support online embedding-based retrieval, we generate embeddings for graph node values using pre-trained text-embedding models like BERT \cite{devlin2018bert} and E5 \cite{wang2022text}, specifically targeting nodes for text-rich sections such as \texttt{"issue summary"}, \texttt{"issue description"}, and \texttt{"steps to reproduce"}, etc. These embeddings are then stored in a vector database (for instance, QDrant \cite{qdrant2024})). For most cases the text-length within each node can meet the text-embedding model's context length constraints, but for certain lengthy texts,we can safely divide the text into smaller chunks for individual embedding without worrying about quality since the text all belong to the same section.


\subsection{Retrieval and Question Answering}
\subsubsection{Query Entity Identification and Intent Detection}
\label{subsubsection:query_entity_identification_and_intent_detection}
In this step, we extract the named entities \(\mathcal{P}\) of type \(\text{Map}(\mathcal{N} \rightarrow \mathcal{V})\) and the query intent set \(\mathcal{I}\) from each user query \(q\). The method involves parsing each query \(q\) into a key-value pair, where each key \(n\), mentioned within the query, corresponds to an element in the graph template \(\mathcal{T}_{\text{template}}\), and the value \(v\) represents the information extracted from the query. Concurrently, the query intents \(\mathcal{I}\) include the entities mentioned in the graph template \(\mathcal{T}_{\text{template}}\) that the query aims to address. We leverage LLM with a suitable prompt in this parsing process. For instance, given the query \(q=\) \textit{"How to reproduce the login issue where a user can't log in to LinkedIn?"}, the extracted entity is \(\mathcal{P}\) = Map(\texttt{"issue summary"} $\rightarrow$ "login issue", \texttt{"issue description"} $\rightarrow$ "user can’t log in to LinkedIn"), and the intent set is \(\mathcal{I}\)=Set(\texttt{"fix solution"}). This method demonstrates notable flexibility in accommodating varied query formulations by leveraging the LLM's extensive understanding and interpretive capabilities.
$$ P, I = \text{LLM}(q, T_{\text{template}}, \text{prompt}) $$

\subsubsection{Embedding-based Retrieval of Sub-graphs}
\label{subsubsection:embedding-based_retrieval_of_sub-graphs}
Our method extracts pertinent sub-graphs from the knowledge graph, aligned with user-provided specifics such as \texttt{"issue description"} and \texttt{"issue summary"}, as well as user intentions like \texttt{"fix solution"}. This process consists of two primary steps: EBR-based ticket identification and LLM-driven subgraph extraction.

In the \textbf{EBR-based ticket identification step}, the top \(K_{\text{ticket}}\) most relevant historical issue tickets are pinpointed by harnessing the named entity set \(\mathcal{P}\) derived from user queries. For each entity pair \((k,v) \in \mathcal{P}\), cosine similarity is computed between the entity value \(v\) and all graph nodes \(n\) corresponding to section \(k\) via pre-trained text embeddings. Aggregating these node-level scores to ticket-level by summing contributions from nodes belonging to the same ticket, we rank and select the top \(K_{\text{ticket}}\) tickets. This method presupposes that the occurrence of multiple query entities is indicative of pertinent links, thus improving retrieval precision.
\[S_{T_i} = \sum_{(k,v) \in \mathcal{P}} \left[ \sum_{n \in T_i} \mathbb{I}\{n.\text{sec} = k\} \cdot \cos(\text{embed}(v), \text{embed}(n.\text{text}))\right]\]

In the \textbf{LLM-driven subgraph extraction step}, the system first rephrases the original user query \(q\) to include the retrieved ticket ID; the modified query \(q'\) is then translated into a graph database language, such as \texttt{Cypher} for \texttt{Neo4j} for question answering. For instance, from the initial query \(q=\)\textit{"how to reproduce the issue where user saw 'csv upload error in updating user email' with major priority due to a data issue"}, the query is reformulated to \textit{"how to reproduce 'ENT-22970'} and thereafter transposed into the Cypher query \texttt{MATCH (j:Ticket \{ticket\_ID: 'ENT-22970'\}) -[:HAS\_DESCRIPTION]-> (description:Description) \\
-[:HAS\_STEPS\_TO\_REPRODUCE]->} \texttt{(steps\_to\_reproduce:\\ StepsToReproduce)} \texttt{RETURN steps\_to\_reproduce.value}. It is noteworthy that the LLM-driven query formulation is sufficiently versatile to retrieve information across subgraphs, whether they originate from the same tree or distinct trees within the knowledge graph.

\subsubsection{Answer Generation}

Answers are synthesized by correlating retrieved data from Section \ref{subsubsection:embedding-based_retrieval_of_sub-graphs} with the initial query. The LLM serves as a decoder to formulate responses to user inquiries given the retrieved information. For robust online serving, if query execution encounters issues, a fallback mechanism reverts to a baseline text-based retrieval method

\section{Experiment}
\label{section: experiment}

\subsection{Experiment Design}
Our evaluation employed a curated "golden" dataset comprising typical queries, support tickets, and their authoritative solutions. The control group operated with conventional text-based EBR, while the experimental group applied the methodology outlined in this study. For both groups, we utilized the same LLM, specifically GPT-4 \cite{achiam2023gpt}, and the same embedding model, E5 \cite{wang2022text}. We measured retrieval efficacy using Mean Reciprocal Rank (MRR), recall@K, and NDCG@K. MRR gauges the average inverse rank of the initial correct response, recall@K determines the likelihood of a relevant item's appearance within the top K selections, and NDCG@K appraises the rank quality by considering both position and pertinence of items. For question-answering performance, we juxtaposed the "golden" solutions against the generated responses, utilizing metrics such as BLEU \cite{papineni2002bleu}, ROUGE \cite{lin2004rouge}, and METEOR \cite{banerjee2005meteor} scores.

\subsection{Result and Analysis}
The retrieval and question-answering performances are presented in Table \ref{tab:results-retrieval} and Table \ref{tab:results-question-answering}, respectively. Across all metrics, our method demonstrates consistent improvements. Notably, it surpasses the baseline by 77.6\% in MRR and by 0.32 in BLEU score, substantiating its superior retrieval efficacy and question-answering accuracy.

\begin{table}[htbp]
\centering
\caption{Retrieval Performance}
\label{tab:results-retrieval}
\begin{tabular}{l S[table-format=1.3] S[table-format=1.3] S[table-format=1.3] S[table-format=1.3] S[table-format=1.3]}
\toprule
& \multicolumn{1}{c}{MRR} 
& \multicolumn{2}{c}{Recall@K} 
& \multicolumn{2}{c}{NDCG@K} \\
& {} & {K=1} & {K=3} & {K=1} & {K=3} \\
\midrule
Baseline    & 0.522 & 0.400 & 0.640 & 0.400 & 0.520 \\
Experiment  & \textbf{0.927} & \textbf{0.860}  & \textbf{1.000} &  \textbf{0.860} & \textbf{0.946} \\
\bottomrule
\end{tabular}
\end{table}

\begin{table}[htbp]
\centering
\caption{Question Answering Performance}
\label{tab:results-question-answering}
\begin{tabular}{
    l
    S[table-format=1.3]
    S[table-format=1.3]
    S[table-format=1.3]
}
\toprule
{} & \text{BLEU} & \text{METEOR} & \text{ROUGE} \\
\midrule
Baseline & 0.057 & 0.279 & 0.183 \\
Experiment & \textbf{0.377} & \textbf{0.613}  & \textbf{0.546} \\
\bottomrule
\end{tabular}
\end{table}

\section{Production Use Case}
We deployed our method within LinkedIn's customer service team, covering multiple product lines. The team was split randomly into two groups: one used our system, while the other stuck to traditional manual methods. As shown in Table \ref{tab:customer-support-performance}, the group using our system achieved significant gains, reducing the median resolution time per issue by 28.6\%. This highlights our system's effectiveness in enhancing customer service efficiency.

\begin{table}[htbp]
\centering
\caption{Customer Support Issue Resolution Time}
\label{tab:customer-support-performance}
\begin{tabular}{
    l
    l
    l
    l
}
\toprule
{Group} & {Mean} & {P50} & {P90} \\
\midrule
Tool Not Used & 40 Hours & 7 Hours & 87 Hours \\
Tool Used & \bfseries 15 hours & \bfseries 5 hours & \bfseries 47 hours \\
\bottomrule
\end{tabular}
\end{table}

\section{Conclusions and Future Work}
In conclusion, our research significantly advances automated question answering systems for customer service. Integrating retrieval augmented generation (RAG) with a knowledge graph (KG) has improved retrieval and answering metrics, and overall service effectiveness. Future work will focus on: developing an automated mechanism for extracting graph templates, enhancing system adaptability; investigating dynamic updates to the knowledge graph based on user queries to improve real-time responsiveness; and exploring the system's applicability in other contexts beyond customer service.

\section{COMPANY PORTRAIT}
 
 \textbf{About LinkedIn}: Founded in 2003, LinkedIn connects the world’s professionals to make them more productive and successful. With more than 1 billion members worldwide, including executives from every Fortune 500 company, LinkedIn is the world’s largest professional network. The company has a diversified business model with revenue coming from Talent Solutions, Marketing Solutions, Sales Solutions and Premium Subscriptions products. Headquartered in Silicon Valley, LinkedIn has offices across the globe. Please visit \href{https://www.linkedin.com/company/linkedin/about/}{https://www.linkedin.com/company/linkedin/about/} for more information.

	\section{PRESENTER BIO}
\textbf{Zhentao Xu} is a Senior Software Engineer at LinkedIn. He received his M.S. in Robotics and B.S. in Electrical Engineering and Computer Science (EECS) from University of Michigan. His research interests lie in large language models and natural language generation.

\bibliographystyle{ACM-Reference-Format}
\balance
\bibliography{main}

\appendix

\end{document}